%% file: entropy-34941-english-layout.tex
\documentclass[journal,article,accept,moreauthors,12pt,a4paper]{mdpi} 

\usepackage{graphicx}
\usepackage{amsfonts}
\usepackage{amssymb}
\usepackage{amsmath}
\usepackage{amsthm,soul}
\lastpage{x}
\doinum{10.3390/------}
\pubvolume{15}
\pubyear{2013}
\history{Received: 24 July 2013; in revised form: 14 October 2013 / Accepted: 14 October 2013 / \\
Published: xx}

\pdfoutput=1
\Title{A Kernel-Based Calculation of Information on a Metric Space}

\Author{R. Joshua Tobin $^{1}$and Conor J. Houghton $^{2,}$*}

\address{%
$^{1}$ School of Mathematics, Trinity College Dublin, Dublin 2, Ireland\\
$^{2}$ Department of Computer Science, University of Bristol, Merchant Venturers Building, Woodland Road, Bristol BS8 1UB, England}

\corres{E-Mail: conor.houghton@bristol.ac.uk; \linebreak Tel.: +44-117-954-5140.}

\abstract{
Kernel density estimation is a technique for approximating probability
distributions. Here, it is applied to the calculation of mutual
information on a metric space. This is motivated by the problem in
neuroscience of calculating the mutual information between stimuli and
spiking responses; the space of these responses is a metric space. It
is shown that kernel density estimation on a metric space resembles the
$k$-nearest-neighbor approach. This approach is applied to a toy
dataset designed to mimic electrophysiological~data.
}


\begin{document}
\vspace{-12pt}
\section{Introduction}

This paper is concerned with the calculation of mutual information for
spike trains using the data that are available in a typical \emph{in
 vivo} electrophysiology experiment in the sensory system. It uses a
kernel-based estimation of probability distributions.

In particular, this paper is concerned with computing the mutual
information, $I(R;S)$, between two random variables, $R$ and $S$. The
motivating neuroscience example is a typical sensory pathway
electrophysiology experiment in which the corpus of sensory stimuli
are presented over multiple trials, so there is a set of recorded
responses for each of a number of stimuli. The stimuli are drawn from
a discrete space, the corpus, but the responses are spike trains. The
space of spike trains is peculiar; locally, it is like a smooth
manifold, with the spike times behaving like coordinates; but, globally,
it is foliated into subspaces, each with a different number of
spikes. The space of spike trains does, however, have a metric. As
such, $S$ takes values in a discrete set, $\cal{S}$, and models the
stimulus, and $R$ takes values in a metric space, $\cal{R}$, and
models the response.

$\cal{R}$ is not a discrete space, and so, to calculate the mutual
information between $S$ and $R$, it is necessary to either discretize
$\cal{R}$ or to use differential mutual information. In the
application of information theory to electrophysiological data, it is
common to take the former route and discretize the data. Here, the
latter alternative is chosen, and the differential mutual information
is estimated.

The mutual information between two random variables, $R$ and $S$, is a
measure of the average amount of information that is gained about $S$
from knowing the value of $R$. With $S$, a discrete random variable
taking values in $\mathcal{S}$ and $R$, a continuous random variable, the mutual
information is:
\begin{linenomath}
\begin{equation} \label{mi_c}
 I(R;S) = \displaystyle \sum_{s\in\mathcal{S}}\int_\mathcal{R} p(r,s) \log_2 \frac{p(r,s)}{p(r)p(s)} dr
\end{equation}
\end{linenomath}
where $dr$ is the measure on $\mathcal{R}$: computing the differential
mutual information between $R$ and $S$ requires integration over
$\cal{R}$. Integration requires a measure, and when there are
coordinates on a space, it is common to use the integration measure
derived from these coordinates. 

The space of spike trains has no system of coordinates, and so, there is
no coordinate-based measure. This does not mean that the space has no
measure. As a sample space, it has an intrinsic measure corresponding
to the probability distribution; thus, there is a measure, just not
one derived from coordinates. The probability of an event occurring in
a region of sample space gives a volume for that region. In other
words, the volume of a region, $\mathcal{D}$, can be identified with
$P(\mathbf{x}\in \mathcal{D})$. This is the measure that will be used
throughout this paper; it does not rely on coordinates, and so, can be
applied to the case of interest here.

Of course, in practice, the probability density is not usually known
on the space of spike trains, but $P(\mathbf{x}\in \mathcal{D})$ can
be estimated from the set of experimental data. A Monte-Carlo-like
approach is used: the volume of a region is estimated by counting the
fraction of data points that lie within it:
\begin{linenomath}
\begin{equation}
\mbox{vol}(\mathcal{D})=P(\mathbf{x}\in \mathcal{D})\approx \frac{\mbox{number of data points in }\mathcal{D}}{\mbox{total number of points}}
\end{equation}
\end{linenomath}
This is exploited in this paper to estimate the volume of square
kernels, making it possible to estimate conditional probabilities
using kernel density estimation.

The classical approach to the problem of estimating $I(R;S)$ is to map
the spike trains to binary words using temporal binning
\cite{DeRuyterVanSteveninckEtAl1997a,BialekEtAl1998a}, giving a
histogram approximation for $p(r,s)$. This approach is very
successful, particularly when supplemented with a strategically chosen
prior distribution for the underlying probability distribution of
words
\cite{NemenmanBialekDeRuyterVanSteveninck2004a,NemenmanEtAl2007a}. This
is sometimes called the plug-in method, and that term is adopted here.
One advantage of the plug-in method is that the mutual information it
calculates is correct in the limit: in the limit of an infinite amount
of data and an infinitesimal bin size, it gives the differential mutual
information.
 
Nonetheless, it is interesting to consider other approaches, and in
this spirit, an alternate approach is presented here. This new method
exploits the inherent metric structure of the space of spike trains,
it is very natural and gives an easily implemented algorithm, which is 
accurate on comparatively small~datasets.

\section{Methods}

This section describes the proposed method for calculating mutual
information. Roughly, the conditional probability is approximated
using kernel density estimation and, by using the unconditioned
probability distribution as a measure, integration is approximated by
the Monte-Carlo method of summing over data points.

Since this is a kernel-based approach, a review of kernel density
estimation is given in Section 2.1. This also serves to establish
notation. The two key steps used to derive the kernel-based estimate
are a change of measure and a Monte-Carlo estimate. The change of
measure, described in Section~2.2, permits the estimation of
probabilities by a simple Monte-Carlo method. The new measure also
simplifies the calculation of $I(R;S)$, resulting in a formula
involving a single conditional distribution. This conditional
distribution is estimated using a Monte-Carlo estimate in Section 2.3.

\subsection{Kernel Density Estimation}

The non-parametric kernel density estimation (KDE) method
\cite{Rosenblatt1956a,Parzen1962a, Silverman1986a} is an approach to
estimating probability densities. In KDE, a probability density is
estimated by filtering the data with a kernel. This kernel is
normalized with an integral of one and is usually symmetric and
localized. For an $n$-dimensional distribution with outcome vectors
$\{\mathbf{x}_1, \mathbf{x}_2,\ldots, \mathbf{x}_m\}$ and a kernel,
$k(\mathbf{x})$, the estimated distribution is usually written:
\begin{linenomath}
\begin{equation}
\tilde{p}(\mathbf{x})=\frac{1}{m}\sum_i k(\mathbf{x}-\mathbf{x}_i)
\end{equation}
\end{linenomath}
where, because the argument is $\mathbf{x}-\mathbf{x}_i$, there is a
copy of the kernel centered at each data point. In fact, this relies
on the vector-space structure of $n$-dimensional space; in the
application considered here, a more general notation is required, with
$k(\mathbf{x};\mathbf{y})$ denoting the value at $\mathbf{x}$ of the
kernel when it is centered on $\mathbf{y}$. In this situation, the
estimate becomes:
\begin{linenomath}
\begin{equation}
\tilde{p}(\mathbf{x})=\frac{1}{m}\sum_i k(\mathbf{x} ; \mathbf{x}_i)
\end{equation}
\end{linenomath}

The square kernel is a common choice. For a vector space, this is:
\begin{linenomath}
\begin{equation}
k(\mathbf{x};\mathbf{y})=\left\{\begin{array}{ll} \frac{1}{V}&\|\mathbf{x} - \mathbf{y}\|<1 \\0&\mbox{otherwise}\end{array}\right.
\end{equation}
\end{linenomath}
where $V$ is chosen, so that the kernel integrates to one. The kernel is usually
scaled to give it a bandwidth:
\begin{linenomath}
\begin{equation}
k(\mathbf{x};\mathbf{y},h)=\left\{\begin{array}{ll} \frac{1}{hV}&\|\mathbf{x} - \mathbf{y}\|<h \\0&\mbox{otherwise}\end{array}\right.
\end{equation}
\end{linenomath}
This bandwidth, $h$, specifies the amount of smoothing. The square
kernel is the most straight-forward choice of kernel mathematically,
and so, in the construction presented here, a square kernel is used. 

In the case that will be of interest here, where $\mathbf{x}$ and
$\mathbf{y}$ are not elements of a vector space, the condition
$\|\mathbf{x} - \mathbf{y}\|<h$ must be replaced by
$d(\mathbf{x},\mathbf{y})<h$, where $d(\mathbf{x},\mathbf{y})$ is a
metric measuring the distance between $\mathbf{x}$ and
$\mathbf{y}$. Calculating the normalization factor, $V$, is more
difficult, since this requires integration. This problem is discussed
in the next subsection.

\subsection{Change of Measure}

Calculating the differential mutual information using KDE requires
integration, both the integration required by the definition of the
mutual information and the integration needed to normalize the
kernel. As outlined above, these integrals are estimated using a
Monte-Carlo approach; this relies on a change of measure, which is
described in this section.

For definiteness, the notation used here is based on the intended
application to spike trains. The number of stimuli is $n_s$, and each
stimulus is presented for $n_t$ trials. The total number of responses,~$n_r$, is then $n_r=n_sn_t$. Points in the set of stimuli are called
$s$ and in the response space, $r$; the actual data points are
indexed, $r_i$, and $(r_i,s_i)$ is a response-stimulus pair. As above,
the random variables for stimulus and response are $S$ and $R$, whereas
the set of stimuli and the space of responses are denoted by a calligraphic 
$\mathcal{S}$ and $\mathcal{R}$, respectively. It is intended that when
the method is applied, the responses, $r\in\mathcal{R}$, will be spike
trains.

The goal is to calculate the mutual information between the stimulus
and the response. Using the Bayes theorem, this is:
\begin{linenomath}
\begin{equation}
 I(R;S) = \sum_{s\in\mathcal{S}}\int_\mathcal{R} p(r,s) \log_2 \frac{p(r|s)}{p(r)} dr 
\end{equation}
\end{linenomath}
Unlike the differential entropy, the differential mutual information
is invariant under the choice of measure. Typically, differential
information theory is applied to examples where there are coordinates,
$(x_1,x_2,\ldots,x_n)$, on the response space and the measure is given
by $dr=dx_1dx_2\ldots dx_n$. However, here, it is intended to use the
measure provided by the probability distribution, $p(r)$. Thus, for a
region, $\mathcal{D}\subset \mathcal{R}$, the change of measure is:
\begin{linenomath}
\begin{equation}
\mbox{vol}(\mathcal{D})=\int_\mathcal{D} p(r) dr = \int_\mathcal{D} d\beta
\end{equation}
\end{linenomath}
so:
\begin{linenomath}
\begin{equation}
d\beta=p(r)dr
\end{equation}
\end{linenomath}
The new probability density relative to the new measure, $p_\beta(r)$,
is now one:
\begin{linenomath}
\begin{equation}
p_\beta(r)=\frac{p(r)}{d\beta/dr}=1
\end{equation}
\end{linenomath}
Furthermore, since $p(r|s)$ and $p(r)$ are both densities,
$p(r|s)/p(r)$ is invariant under a change of measure~and:
\begin{linenomath}
\begin{equation}
 I(R;S) = \sum_{s}\int_\mathcal{R} p_\beta(r,s) \log_2
 \frac{p_\beta(r|s)}{p_\beta(r)} d\beta =\sum_{s}\int_\mathcal{R} p_\beta(r,s) \log_2 p_\beta(r|s) d\beta
\end{equation}
\end{linenomath}
where, again, $p_\beta(r,s)$ and $p_\beta(r|s)$ are the values of the
densities, $p(r,s)$ and $p(r|s)$, after the change of~measure.

The expected value of any function, $f(R,S)$, of random variables, $R$
and $S$, is:
\begin{linenomath}
\begin{equation}
\langle f\rangle = \sum_{s\in\mathcal{S}}\int_\mathcal{R} p_\beta(r,s) f(r,s) d\beta
\end{equation}
\end{linenomath}
and this can be estimated on a set of outcomes, $\{(r_i,s_i)\}$, as:
\begin{linenomath}
\begin{equation}
\langle f\rangle \approx \frac{1}{n_r}\sum_{i} f(r_i,s_i)
\end{equation}
\end{linenomath}
For the mutual information, this gives:
\begin{linenomath}
\begin{equation}
 I(R;S) \approx \frac{1}{n_r}\sum_{i} \log_2{p_\beta(r_i|s_i)}
\end{equation}
\end{linenomath}
Now, an estimate for $p_\beta(r_i|s_i)$ is needed; this is approximated
using KDE.

\subsection{A Monte-Carlo Estimate}

One advantage to using $d\beta$ as the measure is that $p_\beta(r)=1$,
and this simplifies the expression for $I(R;S)$. However, the most
significant advantage is that under this new measure, volumes can be
estimated by simply counting data points. This is used to normalize
the kernel. It is useful to define the support of a function: if
$f(r)$ is a function, then the support of $f(r)$, supp$[f(r)]$, is the
region of its domain where it has a non-zero value:
\begin{linenomath}
\begin{equation}
\mbox{supp}[f(r)]=\{r:f(r) \neq 0\}
\end{equation}
\end{linenomath}
Typically, the size of a square kernel is specified by the radius of
the support. Here, however, it is specified by volume. In a vector
space where the volume measure is derived from the coordinates, there
is a simple formula relating radius and volume. That is not the case
here, and specifying the size of a kernel by volume is not equivalent
to specifying it by radius. Choosing the volume over the radius
simplifies subsequent calculations and, also, has the advantage that the
size of the kernel is related to the number of data points. This also
means that the radius of the kernel varies across $\mathcal{R}$.

The term, bandwidth, will be used to describe the size of the kernel,
even though here, the bandwidth is a volume, rather than a radius. Since
$d\beta$ is a probability measure, all volumes are between zero and
one. Let $h$ be a bandwidth in this range. If $k(r';r,h)$ is the value
at $r'$ of a square kernel with bandwidth $h$ centered on $r$, the
support will be denoted as $\mathcal{S}(r;h)$:
\begin{linenomath}
\begin{equation}
\mathcal{S}(r;h)=\mbox{supp}[k(r';r,h)]
\end{equation}
\end{linenomath}
and the volume of the support of the kernel is vol$[\mathcal{S}(r;h)]$. The 
value of the integral is set at one: 
\begin{linenomath}
\begin{equation}
\int_{\mathcal{S}(r;h)} k(r';r,h) d\beta =1
\end{equation}
\end{linenomath}
and so, since the square kernel is being used, $k(r';r,h)$ has a
constant value of $1/\mbox{vol}[\mathcal{S}(r;h)]$ throughout $\mathcal{S}(r;h)$.

Thus, volumes are calculated using the measure, $d\beta$, based on the
probability density. However, this density is unknown, and so, volumes
need to be estimated. As described above, using $d\beta$, the volume
of a region is estimated by the fraction of data points that lie
within it. In other words, the change of measure leads to a
Monte-Carlo approach to calculating the volume of any region. In the
Monte-Carlo calculation, the volume of the support of a kernel is
estimated as the fraction of data points that lie within it. A choice
of convention has to be made between defining the kernel as
containing $\lfloor hn_r \rfloor$ or $\lceil hn_r \rceil$ points, that
is, on whether to round $hn_r$ down or up. The former choice is used, so,
the kernel around a point, $r$, is estimated as the region containing
the nearest $n_h=\lfloor hn_r \rfloor$ points to $r$, including $r$
itself. Thus, the kernel around a point, $r_i$, is defined as:
\begin{linenomath}
\begin{equation}
k(r;r_i,n_h)=\left\{\begin{array}{ll} \frac{1}{n_h}, & r\mbox{ is one of the }n_h\mbox{ closest points to }r_i\\0,&\mbox{otherwise}\end{array}\right.
\end{equation}
\end{linenomath}
and the support, $\mathcal{S}(r_i;n_h)$, has $r_j\in
\mathcal{S}(r_i;n_h)$ if $k(r_j;r_i,n_h)=1/n_h$, or, put another way,
$r_j$ is one of the $n_h$ nearest data points. In practice, rather
than rounding $hn_r$ up or down, the kernel volume in a particular
example can be specified using $n_h$ rather than $h$.

Typically, kernels are balls: regions defined by a constant radius. As
such, the kernel described here makes an implicit assumption about the
isotropic distribution of the data points. However, in the normal
application of KDE, special provision must be made near boundaries,
where the distribution of data points is not isotropic
\cite{Jones1993a}. Here, these cases are dealt with automatically.

Since $p_\beta(r_i|s_i)=n_sp_\beta(r_i,s_i)$, here, the conditional 
distribution, $p_\beta(r_i | s_i)$, is estimated by
first estimating $p_\beta(r_i,s_i)$. As described above, a kernel has
a fixed volume relative to the measure based on $p_\beta(r)$. Here, the
kernel is being used to estimate $p_\beta(r_i,s_i)$:
\begin{linenomath}
\begin{equation}
\tilde{p}_\beta(r_i,s_i)=\frac{c(r_i,s_i;n_h)}{n_h}
\end{equation}
\end{linenomath}
where $c(r_i,s_i;n_h)$ is the number of data points evoked to stimulus
$s_i$ for which $r_i$ is one of the $n_h$ closest~points:
\begin{linenomath}
\begin{equation}
 c(r_i,s_i;n_h)=|\{(r_j,s_i):r_j\in \mathcal{S}(r_i;n_h)\}|
\end{equation}
\end{linenomath}
This gives the estimated mutual information:
\begin{linenomath}
\begin{equation} \label{our_estimate}
 I(R;S) \approx I(R,S;n_h)=\frac{1}{n_r}\sum_{i} \log_2{\frac{n_sc(r_i,s_i;n_h)}{n_h}}
\end{equation}
\end{linenomath}
Remarkably, although this is a KDE estimator, it resembles a $k$-, or,
here, $n_h$-, nearest-neighbors estimator. Basing KDE on the data
available for spike trains appears to lead naturally to nearest
neighbor~estimation.

The formula for $I(R,S;n_h)$ behaves well in the extreme cases. If the
responses to each stimulus are close to each other, but distant from
responses to all other stimuli, then $c(r_i,s_i;n_h)=n_h$ for all
stimulus-response pairs $(r_i,s_i)$. That is, for each data point,
all nearby data points are from the same stimulus. This means that
the estimate will be:
\begin{linenomath}
\begin{equation}
 I(R,S;n_h) = \log_2{n_s}
\end{equation}
\end{linenomath}
This is the correct value, because, in this case, the response
completely determines the stimulus, and so, the mutual information is
exactly the entropy of the stimulus. On the other hand, if the
responses to each stimulus have the same distribution, then
$c(r_i,s_i;n_h)/n_h\approx 1/n_s$, so the estimated mutual information
will be close to zero. This is again the correct value, because in
this case, the response is independent of the stimulus.

\section{Results}

As a test, this method has been applied to a toy dataset modelled on
the behavior of real spike trains. It is important that the method is
applied to toy data that resemble the data type, electrophysiological
data, on which the method is intended to perform well. As such, the toy
model is selected to mimic the behavior of sets of spike trains. The
formula derived above acts on the matrix of inter-data-point distances,
rather than the points themselves, and so, the dataset is designed to
match the distance distribution observed in real spike trains
\cite{GillespieHoughton2009a}. The test dataset is also designed to
present a stiff challenge to any algorithm for estimating information.

The toy data are produced by varying the components of one of a set of
source vectors. More precisely, to produce a test dataset, a variance,
$\sigma^2$, is chosen uniformly from $[0,1]$, and $n_s$ sources are
chosen uniformly in a $n_d$-dimensional box centered at the origin
with unit sides parallel to the Cartesian axes. Thus, the sources are
all $n_d$-dimensional vectors. The data points are also
$n_d$-dimensional vectors; they are generated by drawing each
component from a normal distribution about the corresponding component
of the source. Thus, data points with a source
$\mathbf{s}=(s_1,s_2,\ldots,s_{n_d})$ are chosen as
\mbox{$\mathbf{r}=(r_1,r_2,\ldots,r_{n_d})$}, where the $r_i$ are all drawn
from normal distributions with variance $\sigma^2$ centered at the
corresponding $s_i$:
\begin{linenomath}
\begin{equation}
r_i\sim \mathcal{N}(s_i,\sigma^2)
\end{equation}
\end{linenomath}
$n_t$ data points are chosen for each source, giving $n_r=n_sn_t$ data
points in all. 

Each test uses 200 different datasets; random pruning is used to
ensure that the values of mutual information are evenly distributed over
the whole range from zero to $\log_2{n_s}$; otherwise, there tends to
be an excess of datasets with a low value. The true mutual
information is calculated using a Monte-Carlo estimate sampled over
10,000 points. The actual probability distributions are known: the
probability of finding a point $\mathbf{r}$ generated by a source,
$\mathbf{s}$, depends only on the distance $d=|\mathbf{r}-\mathbf{s}|$
and is given by the $\chi$-distribution:
\begin{linenomath}
\begin{equation}
p(d)=\frac{2^{1-n_d/2}}{\Gamma(n_d/2)}\left(\frac{d}{\sigma}\right)^{n_d-1}e^{-d^2/2\sigma^2}
\end{equation}
\end{linenomath}

There is a bias in estimating the mutual information, in fact, bias is
common to any approach to estimating mutual information
\cite{Paninski2003a}. The problem of reducing bias, or defining the
mutual information, so that the amount of bias is low, is well studied
and has produced a number of sophisticated approaches~\mbox{\cite{Paninski2003a, TrevesPanzeri1995a,PanzeriTreves1996a,
 NemenmanEtAl2007a, PanzeriEtAl2007a,Montemurro2007a}}. One of these,
quadratic estimation, thanks to
\cite{TrevesPanzeri1995a,PanzeriEtAl2007a}, is adapted to the current
situation. Basically, it is assumed that for large numbers of data
points, $n_t$, the estimated information, $\tilde{I}(R;S)$, is related to
the true mutual information $I(R;S)$ by:
\begin{linenomath}
\begin{equation}
\tilde{I}(R;S)=I(R;S)+\frac{A}{n_t}+\frac{B}{n_t^2}+O(1/n_t^3)
\end{equation}
\end{linenomath}
This asymptotic expansion is well-motivated in the case of the plug-in
approach to spike train information~\cite{Miller1955a,Carlton1969a,TrevesPanzeri1995a,Victor2000a,
 Paninski2003a}, and since the sources of bias are presumably
similar, it is assumed the same expansion applies. In fact, this
assumption is supported by plots of $I(R,S;n_h)$ against $n_t$. To
extract $I(R;S)$, the estimate, $I(R,S;n_h)$, is calculated for $\lambda
n_r$ with $\lambda$ taking values from 0.1 to one in 0.1
increments. Least squares fit is used to estimate $I(R;S)$ from these
ten values.

The new method works well on these toy data. It is compared to a
histogram approach, where the $n_d$-dimensional space is discretized
into bins and counting is used to estimate the probability of each
bin. This is an analog of the plug-in method, and the same quadratic
estimation technique is used to reduce~bias.

\begin{figure}[H]
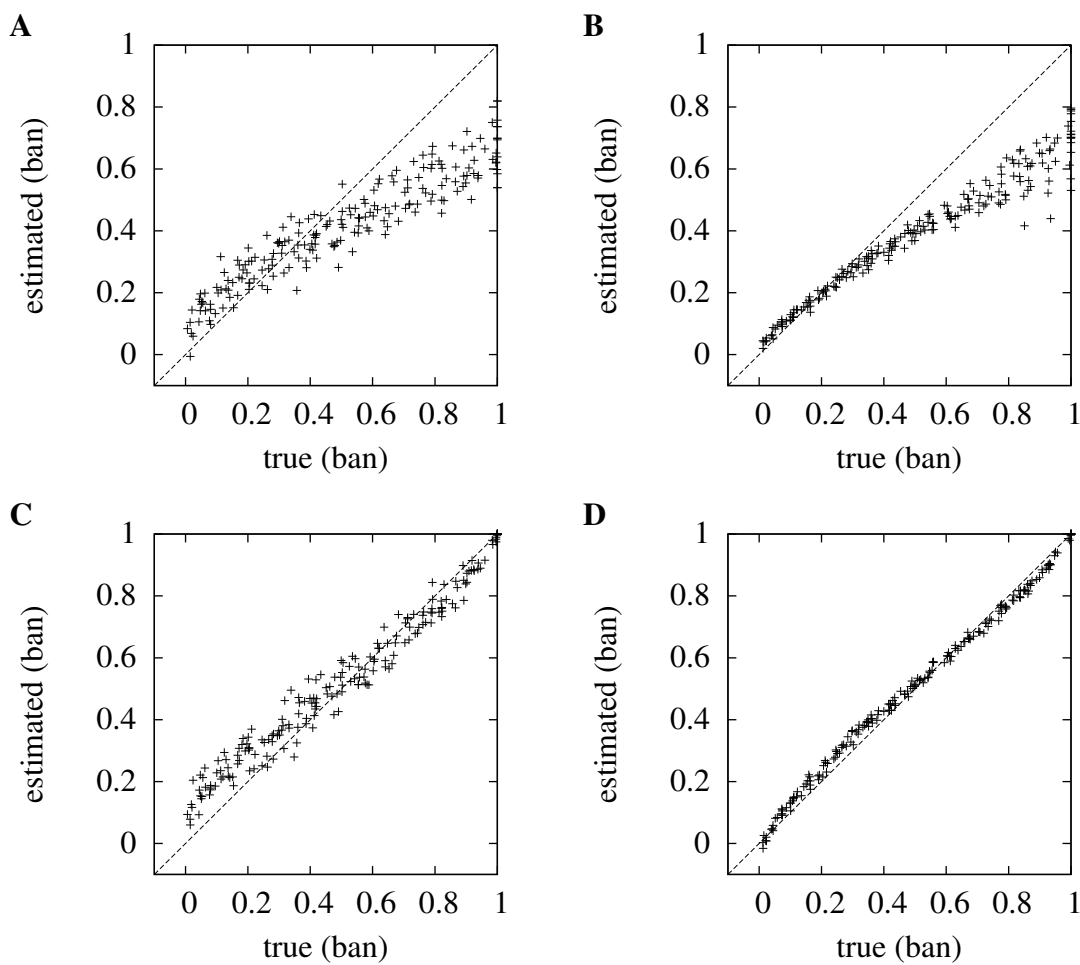

\begin{center}
\include{plot_run1-4}
\end{center}
\vskip -1cm
\caption{Comparing kernel density estimation (KDE) to the histogram method for ten sources,
 $n_s=10$, and three dimensions, $n_d=3$. In each case, the true
 information is plotted against the estimated information; the line,
 $y = x$, which represents perfect estimation, is plotted for
 clarity. For convenience, the mutual information has been
 normalized, so in each case, the value plotted is the estimate of
 $I(R;S)/\log_2{n_s}$, with a maximum value of one; in the cases
 plotted here, that means the information is measured in
 ban. \textbf{A} and \textbf{B} show the distribution for the
 histogram method for $n_t=10$ and $n_t=200$; \textbf{C} and
 \textbf{D} show the kernel~method. \label{fig_comparisons_10_3}}
\end{figure}

In Figure~\ref{fig_comparisons_10_3}, the new method is compared to the
histogram method when $n_s=10$ and $n_d=3$, and for both low and high
numbers of trials, $n_t=10$ and $n_t=200$. For the histogram method,
the optimum discretization width is used. This optimal width is
large, $h=5$ in each case; this roughly corresponds to a different bin
for each octant of the three-dimensional space containing the data. In
the new method, the bandwidth is not optimized on a case by case basis;
instead, the kernel bandwidth, $n_h$, is chosen as being equal to the
number of trials, $n_t$. It can be seen that the new method is better
at estimating the information: for $n_t=10$, it has an average absolute
error of $0.189$ bits, compared to $0.481$ bits for the histogram
method; for $n_t=200$, the average absolute error is $0.083$ bits,
compared to $0.442$ bits for the histogram approach.

\begin{figure}[H]
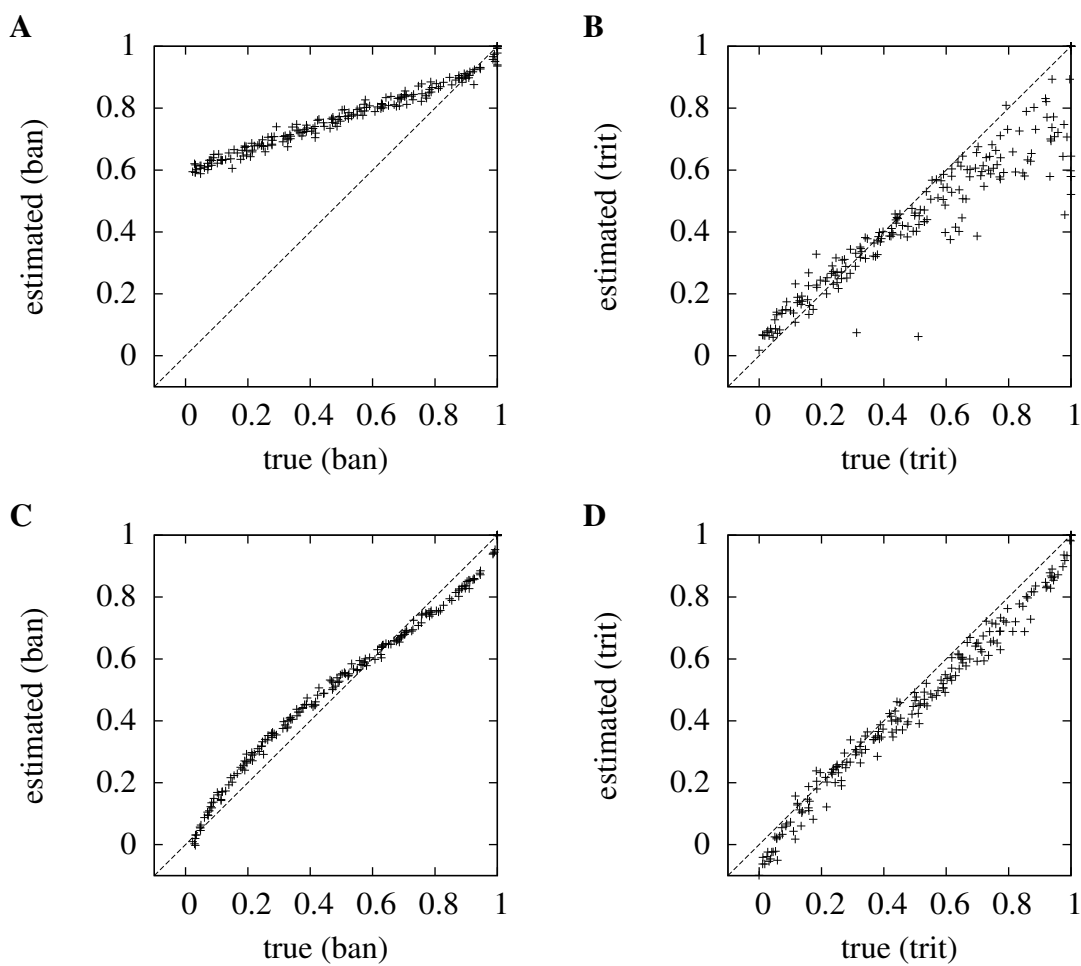

\begin{center}
\include{plot_run7-8}
\end{center}
\vskip -1cm
\caption{Comparing the KDE to the histogram method for high and low
 numbers of sources and dimensions. The true information is plotted
 against the estimated information; in \textbf{A} and~\textbf{C},
 $n_s=10$ and $n_d=10$; in \textbf{B} and \textbf{D}, $n_s=3$ and
 $n_d=3$. The top row, \textbf{A} and \textbf{B}, is for the
 histogram method, the bottom row, \textbf{C} and \textbf{D}, is for
 the kernel method. As before, the normalized information,
 $I(S;R)/\log_2{n_s}$, is plotted. So, for $n_s=10$, the information is
 in ban, for $n_s=3$, in trit, and in each case, the maximum
 mutual information is one. $n_r=200$ for all
 graphs. \label{fig_comparisons_other}}
\end{figure}

In Figure~\ref{fig_comparisons_other}, the histogram and kernel methods
are compared for $n_s=10$ and $n_d=10$ and for $n_s=3$ and $n_d=3$;
the number of trials is $n_t=200$ in each case. The kernel
method outperforms the histogram method. When $n_s=10$ and $n_d=10$,
the average absolute error for the kernel method is 0.139 bits,
compared to 0.876 bits for the histogram method; for $n_s=3$ and
$n_d=3$, its average absolute error is 0.076 bits compared to 0.141 bits for
the histogram. Furthermore, the errors for the kernel method are less
clearly modulated by the actual information, which makes the method
less prone to producing misleading results.

\section{Discussion}

Although the actual method presented here is very different, it was
inspired in part by the transmitted information method for calculating
mutual information using metric-based clustering described in
\cite{VictorPurpura1996a} and by the novel approach introduced in
\cite{BrasseletJohanssonArleo2011a}, where a kernel-like approach to
mutual information is developed. Another significant motivation was
the interesting technique given in \cite{Victor2002a}, where the
information is estimated by measuring how large a sphere could be
placed around each data point without it touching another data
point. In \cite{Victor2002a}, the actual volume of the sphere is
required, or, rather, the rate the volume changes with diameter. This is
calculated by foliating the space of spike trains into subspaces with
a fixed spike number and interpreting the spike times as
coordinates. This is avoided here by using the Monte-Carlo estimate of
volumes. Finally, the copula construction is related to the approach
described here. In fact, the construction here can be thought of as a
reverse copula construction \cite{CalsaveriniVicente2009a}.

An important part of the derivation of the kernel method is the change
of measure to one based on the distribution. Since the kernel size is
defined using a volume based on this measure, the radius of the kernel
adapts to the density of data points. This is similar to the adaptive
partitioning described, for example, in \cite{DarbellayVajda1999a}.
Like the plug-in method of computing mutual information for spike
trains, adaptive partitioning is a discretization approach. However,
rather than breaking the space into regions of fixed width, the
discrete regions are chosen dynamically, using estimates of the
cumulative distribution, similar to what is proposed here.

One striking aspect of KDE seen here is that it reduces to a $k$th
nearest-neighbor (kNN) estimator. The kNN approach to estimating the
mutual information of variables lying in metric spaces has been
studied directly in \cite{KraskovEtAl2004a}. Rather than using a KDE
of the probability distribution, a Kozachenko-Leonenko estimator
\cite{KozachenkoLeonenko1987a} is used. To estimate $I(X;Y)$, where $X$
and $Y$ are both continuous random variables taking values in
$\mathcal{X}$ and $\mathcal{Y}$, Kozachenko-Leonenko estimates are
calculated for $H(X)$, $H(Y)$ and $H(X,Y)$; by using different values
of $k$ in each space, the terms that would otherwise depend on the
dimension of $\mathcal{X}$ and $\mathcal{Y}$ cancel. 

This approach can be modified to estimate $I(R;S)$, where $S$ is a
discrete random variable. Using the approach described in
\cite{KraskovEtAl2004a} to estimate $H(R)$ and $H(R|S)$ gives:
\begin{linenomath}
 \begin{equation} \label{kraskov}
 I_e(R;S) \approx \digamma(n_k) + \digamma(n_tn_s)-\digamma(n_t) - \displaystyle \frac{1}{n_r} \sum_{i} \digamma[C(r_i,s_i;n_k)] 
 \end{equation}
\end{linenomath} 
where $\digamma(x)$ is the digamma function, $n_k$ is an integer
parameter and $C(r_i,s_i;n_k)$ is similar to $c(r_i, s_i;n_h)$ above.
Whereas $c_k(r_i,s_i;n_h)$ counts the number of responses to $s_i$ for
which $r_i$ is one of the $n_h$ closest data points, $C(r_i,s_i;n_k)$
is computed by first finding the distance, $d$, from $r_i$ to the
$n_k$th nearest spike-train response to stimulus $s_i$; then,
$C(r_i,s_i;n_k)$ counts the number of spike trains, from any stimulus,
that is at most a distance of $d$ from $r_i$. $I_e(R;S)$ is the
mutual information with base $e$, so \mbox{$I(R;S)=I_e(R;S)/\ln{2}$}. During
the derivation of this formula, expressions involving the dimension of
$\mathcal{S}$ appear, but ultimately, they all cancel, leaving an
estimate which can be applied in the case of interest here, where
$\mathcal{S}$ has no dimension. Since the digamma function can be
approximated as:
\begin{linenomath}
\begin{equation}
\digamma(x)\approx\ln{x}-\frac{1}{2x}
\end{equation}
\end{linenomath}
for large $x$, this kNN approach and the kernel method produce very
similar estimates. The similarity between the two formulas, despite
the different routes taken to them, lends credibility to both
estimators.

Other versions of the kernel method can be envisaged. A kernel with a
different shape could be used or the kernel could be defined by the
radius rather than by the volume of the support. The volume of the
support and, therefore, the normalization would then vary from data
point to data point. This volume could be estimated by counting, as it
was here. However, as mentioned above, the volume-based bandwidth has
the advantage that it gives a kernel that is adaptive: the radius
varies as the density of data points changes. Another intriguing
possibility is to investigate if it would be possible to follow~
\cite{Victor2002a} and \cite{KraskovEtAl2004a} more closely than has
been done here and use a Monte-Carlo volume estimate to derive a
Kozachenko and Leonenko estimator. Finally, KDE applied to two
continuous random variables could be used to derive an estimate for
the mutual information between two sets of spike trains or between a
set of spike trains and a non-discrete stimulus, such as position in a
maze.

There is no general, principled approach to choosing bandwidths for
KDE methods. There are heuristic methods, such as cross-validation
\cite{Rudemo1982a,Hall1983a}, but these include implicit assumptions
about how the distribution of the data is itself drawn from a family
of distributions, assumptions that may not apply to a particular
experimental situation. The KDE approach developed here includes a
term analogous to bandwidth, and although a simple choice of this
bandwidth is suggested and gives accurate estimates, the problem of
optimal bandwidth selection will require further study.

Applying the KDE approach to spike trains means it is necessary to
specify a spike train metric~\mbox{\cite{VictorPurpura1996a,VanRossum2001a,HoughtonVictor2009a}}. Although
the metric is only used to arrange points in the order of proximity, the
dependence on a metric does mean that the estimated mutual information
will only include mutual information encoded in features of the spike
train that affect the metric. As described in \cite{Victor2002a}, in
the context of another metric-dependent estimator of mutual
information, this means the mutual information may underestimate the
true mutual information, but it does allow the coding structure of
spike trains to be probed by manipulating the spike train
metrics. 

It is becoming increasingly possible to measure large number spike
trains from large numbers of spike trains simultaneously. There are
metrics for measuring distances between sets of multi-neuron responses~\cite{AronovEtAl2003a,HoughtonSen2008a,KreuzEtAl2013a}, and so, the
approach described here can also be applied to multi-neuronal data.

\acknowledgements{Acknowledgments} 

R. Joshua Tobin is grateful to the Irish
Research Council in Science, Engineering and Technology for financial
support. Conor J. Houghton is grateful to the James S. McDonnell Foundation for
financial support through a Scholar Award in Human Cognition.

\acknowledgements{Conflicts of Interest}

The authors declare no conflicts of interest. 

\bibliographystyle{mdpi}
\makeatletter
\renewcommand\@biblabel[1]{#1. }
\makeatother

\end{document}

%% file: plot_run1-4.tex



\begin{tabular}{ll}
\textbf{A}&\textbf{B}\\
  \setlength{\unitlength}{0.0500bp}%
  \begin{picture}(4040.00,3228.00)%
      \put(949,937){\makebox(0,0)[r]{\strut{} 0}}%
      \put(949,1402){\makebox(0,0)[r]{\strut{} 0.2}}%
      \put(949,1867){\makebox(0,0)[r]{\strut{} 0.4}}%
      \put(949,2332){\makebox(0,0)[r]{\strut{} 0.6}}%
      \put(949,2798){\makebox(0,0)[r]{\strut{} 0.8}}%
      \put(949,3263){\makebox(0,0)[r]{\strut{} 1}}%
      \put(1314,484){\makebox(0,0){\strut{} 0}}%
      \put(1779,484){\makebox(0,0){\strut{} 0.2}}%
      \put(2244,484){\makebox(0,0){\strut{} 0.4}}%
      \put(2709,484){\makebox(0,0){\strut{} 0.6}}%
      \put(3175,484){\makebox(0,0){\strut{} 0.8}}%
      \put(3640,484){\makebox(0,0){\strut{} 1}}%
      \put(179,1983){\rotatebox{-270}{\makebox(0,0){\strut{}estimated (ban)}}}%
      \put(2360,154){\makebox(0,0){\strut{}true (ban)}}
    \put(-500,0){\includegraphics{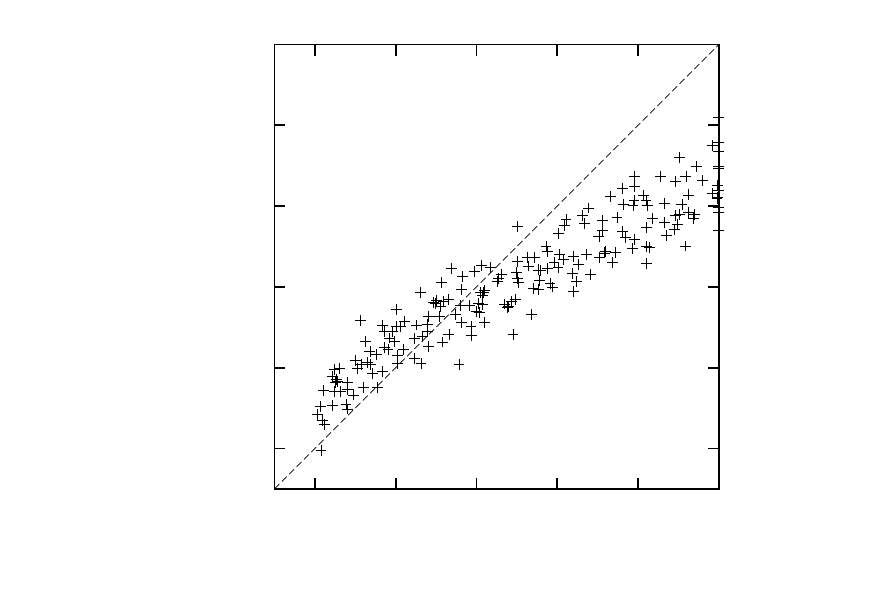}}%
  \end{picture}%
&
  \setlength{\unitlength}{0.0500bp}%
  \begin{picture}(4040.00,3228.00)%
      \put(949,937){\makebox(0,0)[r]{\strut{} 0}}%
      \put(949,1402){\makebox(0,0)[r]{\strut{} 0.2}}%
      \put(949,1867){\makebox(0,0)[r]{\strut{} 0.4}}%
      \put(949,2332){\makebox(0,0)[r]{\strut{} 0.6}}%
      \put(949,2798){\makebox(0,0)[r]{\strut{} 0.8}}%
      \put(949,3263){\makebox(0,0)[r]{\strut{} 1}}%
      \put(1314,484){\makebox(0,0){\strut{} 0}}%
      \put(1779,484){\makebox(0,0){\strut{} 0.2}}%
      \put(2244,484){\makebox(0,0){\strut{} 0.4}}%
      \put(2709,484){\makebox(0,0){\strut{} 0.6}}%
      \put(3175,484){\makebox(0,0){\strut{} 0.8}}%
      \put(3640,484){\makebox(0,0){\strut{} 1}}%
      \put(179,1983){\rotatebox{-270}{\makebox(0,0){\strut{}estimated (ban)}}}%
      \put(2360,154){\makebox(0,0){\strut{}true (ban)}}%
    \put(-500,0){\includegraphics{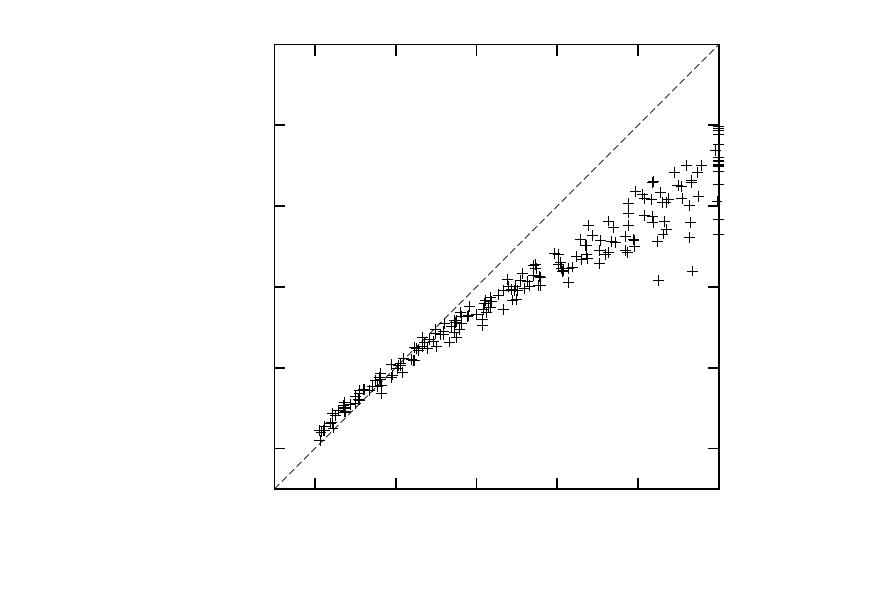}}%
  \end{picture}%
\\
\textbf{C}&\textbf{D}\\
  \setlength{\unitlength}{0.0500bp}%
  \begin{picture}(4040.00,3228.00)%
      \put(949,937){\makebox(0,0)[r]{\strut{} 0}}%
      \put(949,1402){\makebox(0,0)[r]{\strut{} 0.2}}%
      \put(949,1867){\makebox(0,0)[r]{\strut{} 0.4}}%
      \put(949,2332){\makebox(0,0)[r]{\strut{} 0.6}}%
      \put(949,2798){\makebox(0,0)[r]{\strut{} 0.8}}%
      \put(949,3263){\makebox(0,0)[r]{\strut{} 1}}%
      \put(1314,484){\makebox(0,0){\strut{} 0}}%
      \put(1779,484){\makebox(0,0){\strut{} 0.2}}%
      \put(2244,484){\makebox(0,0){\strut{} 0.4}}%
      \put(2709,484){\makebox(0,0){\strut{} 0.6}}%
      \put(3175,484){\makebox(0,0){\strut{} 0.8}}%
      \put(3640,484){\makebox(0,0){\strut{} 1}}%
      \put(179,1983){\rotatebox{-270}{\makebox(0,0){\strut{}estimated (ban)}}}%
      \put(2360,154){\makebox(0,0){\strut{}true (ban)}}%
    \put(-500,0){\includegraphics{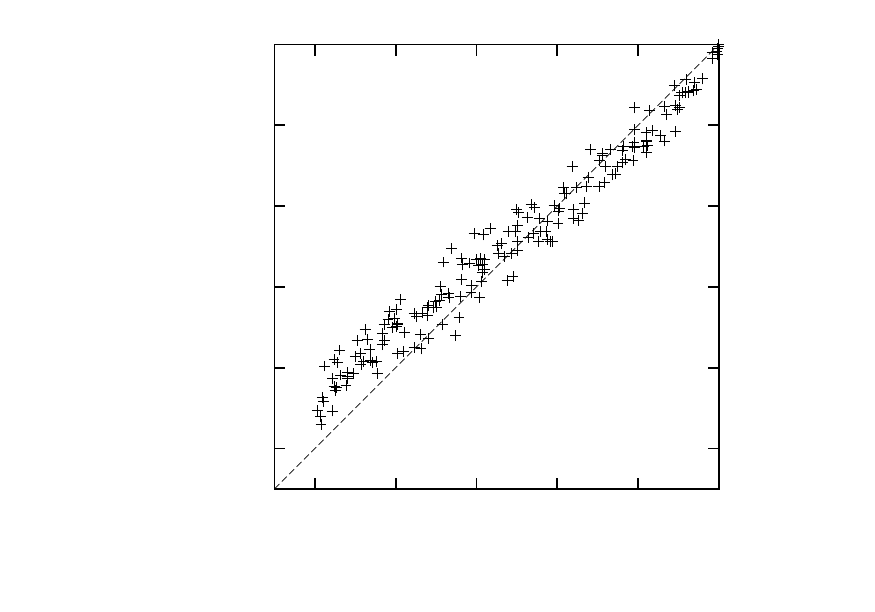}}%
  \end{picture}%
&
  \setlength{\unitlength}{0.0500bp}%
  \begin{picture}(4040.00,3228.00)%
      \put(949,937){\makebox(0,0)[r]{\strut{} 0}}%
      \put(949,1402){\makebox(0,0)[r]{\strut{} 0.2}}%
      \put(949,1867){\makebox(0,0)[r]{\strut{} 0.4}}%
      \put(949,2332){\makebox(0,0)[r]{\strut{} 0.6}}%
      \put(949,2798){\makebox(0,0)[r]{\strut{} 0.8}}%
      \put(949,3263){\makebox(0,0)[r]{\strut{} 1}}%
      \put(1314,484){\makebox(0,0){\strut{} 0}}%
      \put(1779,484){\makebox(0,0){\strut{} 0.2}}%
      \put(2244,484){\makebox(0,0){\strut{} 0.4}}%
      \put(2709,484){\makebox(0,0){\strut{} 0.6}}%
      \put(3175,484){\makebox(0,0){\strut{} 0.8}}%
      \put(3640,484){\makebox(0,0){\strut{} 1}}%
      \put(179,1983){\rotatebox{-270}{\makebox(0,0){\strut{}estimated (ban)}}}%
      \put(2360,154){\makebox(0,0){\strut{}true (ban)}}%
    \put(-500,0){\includegraphics{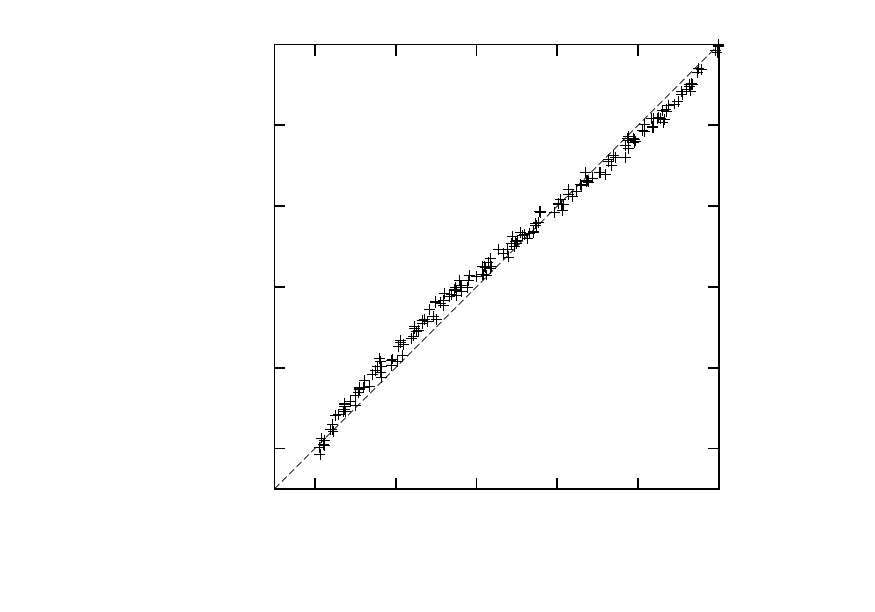}}%
  \end{picture}%
 \end{tabular} 

%% file: plot_run7-8.tex



\begin{tabular}{ll}
\textbf{A}&\textbf{B}\\
  \setlength{\unitlength}{0.0500bp}%
  \begin{picture}(4040.00,3228.00)%
      \put(949,937){\makebox(0,0)[r]{\strut{} 0}}%
      \put(949,1402){\makebox(0,0)[r]{\strut{} 0.2}}%
      \put(949,1867){\makebox(0,0)[r]{\strut{} 0.4}}%
      \put(949,2332){\makebox(0,0)[r]{\strut{} 0.6}}%
      \put(949,2798){\makebox(0,0)[r]{\strut{} 0.8}}%
      \put(949,3263){\makebox(0,0)[r]{\strut{} 1}}%
      \put(1314,484){\makebox(0,0){\strut{} 0}}%
      \put(1779,484){\makebox(0,0){\strut{} 0.2}}%
      \put(2244,484){\makebox(0,0){\strut{} 0.4}}%
      \put(2709,484){\makebox(0,0){\strut{} 0.6}}%
      \put(3175,484){\makebox(0,0){\strut{} 0.8}}%
      \put(3640,484){\makebox(0,0){\strut{} 1}}%
      \put(179,1983){\rotatebox{-270}{\makebox(0,0){\strut{}estimated (ban)}}}%
      \put(2360,154){\makebox(0,0){\strut{}true (ban)}}
    \put(-500,0){\includegraphics{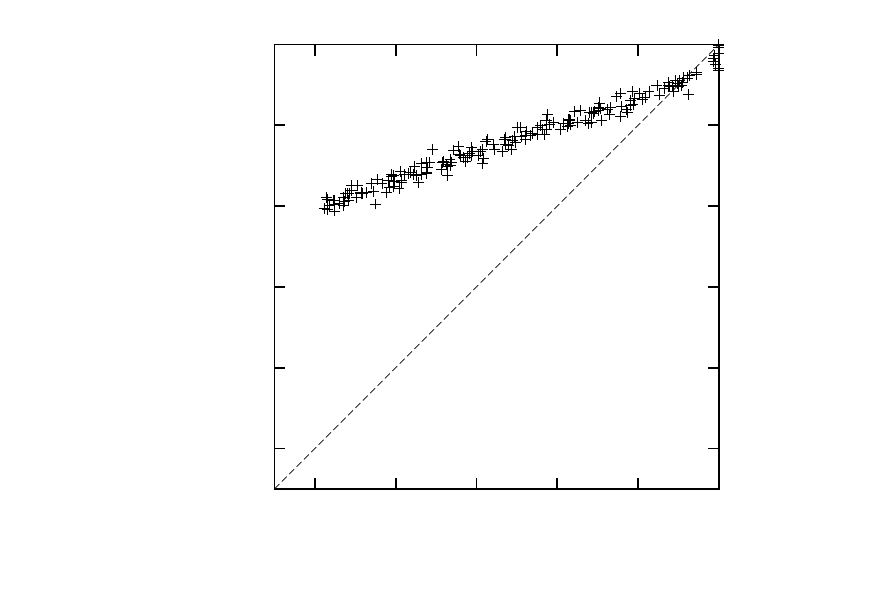}}%
  \end{picture}%
&
  \setlength{\unitlength}{0.0500bp}%
  \begin{picture}(4040.00,3228.00)%
      \put(949,937){\makebox(0,0)[r]{\strut{} 0}}%
      \put(949,1402){\makebox(0,0)[r]{\strut{} 0.2}}%
      \put(949,1867){\makebox(0,0)[r]{\strut{} 0.4}}%
      \put(949,2332){\makebox(0,0)[r]{\strut{} 0.6}}%
      \put(949,2798){\makebox(0,0)[r]{\strut{} 0.8}}%
      \put(949,3263){\makebox(0,0)[r]{\strut{} 1}}%
      \put(1314,484){\makebox(0,0){\strut{} 0}}%
      \put(1779,484){\makebox(0,0){\strut{} 0.2}}%
      \put(2244,484){\makebox(0,0){\strut{} 0.4}}%
      \put(2709,484){\makebox(0,0){\strut{} 0.6}}%
      \put(3175,484){\makebox(0,0){\strut{} 0.8}}%
      \put(3640,484){\makebox(0,0){\strut{} 1}}%
      \put(179,1983){\rotatebox{-270}{\makebox(0,0){\strut{}estimated (trit)}}}%
      \put(2360,154){\makebox(0,0){\strut{}true (trit)}}%
    \put(-500,0){\includegraphics{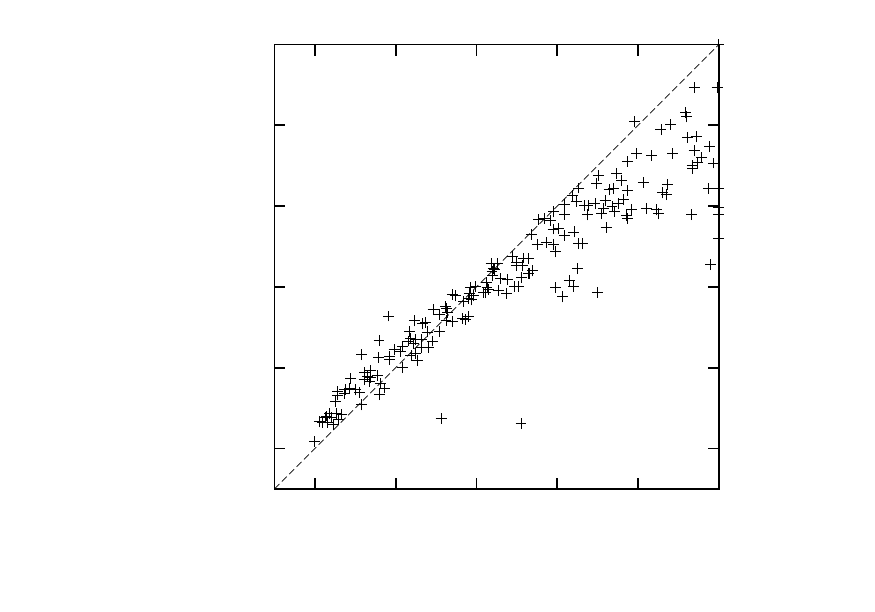}}%
  \end{picture}%
\\
\textbf{C}&\textbf{D}\\
  \setlength{\unitlength}{0.0500bp}%
  \begin{picture}(4040.00,3228.00)%
      \put(949,937){\makebox(0,0)[r]{\strut{} 0}}%
      \put(949,1402){\makebox(0,0)[r]{\strut{} 0.2}}%
      \put(949,1867){\makebox(0,0)[r]{\strut{} 0.4}}%
      \put(949,2332){\makebox(0,0)[r]{\strut{} 0.6}}%
      \put(949,2798){\makebox(0,0)[r]{\strut{} 0.8}}%
      \put(949,3263){\makebox(0,0)[r]{\strut{} 1}}%
      \put(1314,484){\makebox(0,0){\strut{} 0}}%
      \put(1779,484){\makebox(0,0){\strut{} 0.2}}%
      \put(2244,484){\makebox(0,0){\strut{} 0.4}}%
      \put(2709,484){\makebox(0,0){\strut{} 0.6}}%
      \put(3175,484){\makebox(0,0){\strut{} 0.8}}%
      \put(3640,484){\makebox(0,0){\strut{} 1}}%
      \put(179,1983){\rotatebox{-270}{\makebox(0,0){\strut{}estimated (ban)}}}%
      \put(2360,154){\makebox(0,0){\strut{}true (ban)}}%
    \put(-500,0){\includegraphics{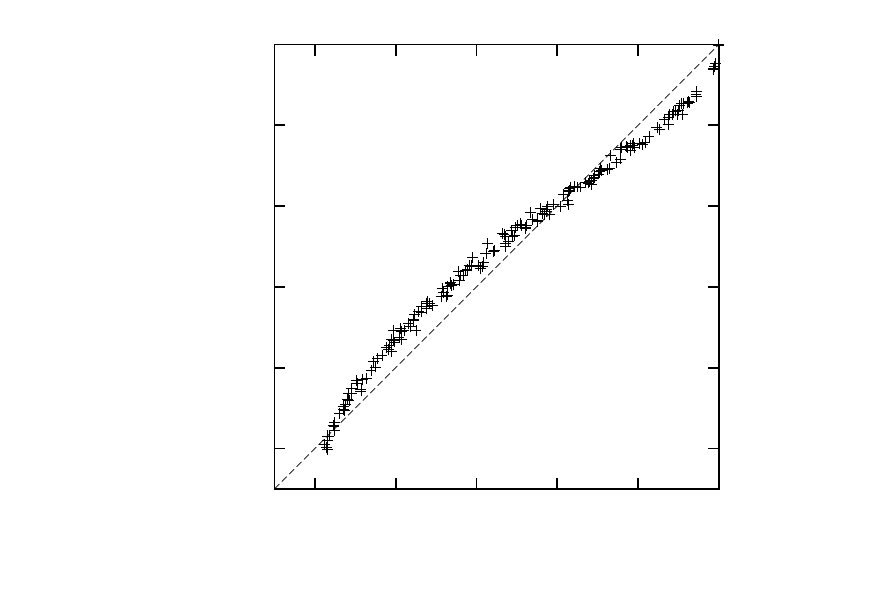}}%
  \end{picture}%
&
  \setlength{\unitlength}{0.0500bp}%
  \begin{picture}(4040.00,3228.00)%
      \put(949,937){\makebox(0,0)[r]{\strut{} 0}}%
      \put(949,1402){\makebox(0,0)[r]{\strut{} 0.2}}%
      \put(949,1867){\makebox(0,0)[r]{\strut{} 0.4}}%
      \put(949,2332){\makebox(0,0)[r]{\strut{} 0.6}}%
      \put(949,2798){\makebox(0,0)[r]{\strut{} 0.8}}%
      \put(949,3263){\makebox(0,0)[r]{\strut{} 1}}%
      \put(1314,484){\makebox(0,0){\strut{} 0}}%
      \put(1779,484){\makebox(0,0){\strut{} 0.2}}%
      \put(2244,484){\makebox(0,0){\strut{} 0.4}}%
      \put(2709,484){\makebox(0,0){\strut{} 0.6}}%
      \put(3175,484){\makebox(0,0){\strut{} 0.8}}%
      \put(3640,484){\makebox(0,0){\strut{} 1}}%
      \put(179,1983){\rotatebox{-270}{\makebox(0,0){\strut{}estimated (trit)}}}%
      \put(2360,154){\makebox(0,0){\strut{}true (trit)}}%
    \put(-500,0){\includegraphics{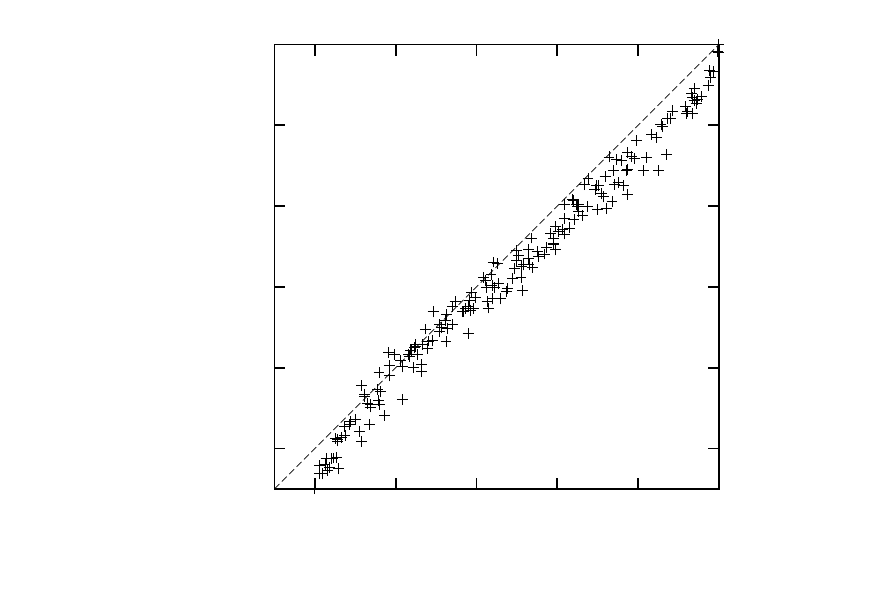}}%
  \end{picture}%
 \end{tabular} 